\documentclass{JHEP}
\usepackage{epsfig}

\usepackage{amsmath}
\usepackage{amssymb}
                                           
\catcode`\@=11
\makeatletter \ifcase\@ptsize \font\teneufm=eufm10
\font\seveneufm=eufm7 \font\fiveeufm=eufm5
\font\teneusm=eusm10 \font\seveneusm=eusm7
\font\fiveeusm=eusm5 \or \font\teneufm=eufm10 scaled
\magstephalf \font\seveneufm=eufm7 \font\fiveeufm=eufm5
\font\teneusm=eusm10 scaled \magstephalf
\font\seveneusm=eusm7 \font\fiveeusm=eusm5 \or
\font\teneufm=eufm10 scaled \magstep1 \font\seveneufm=eufm7
\font\fiveeufm=eufm5 \font\teneusm=eusm10 scaled \magstep1
\font\seveneusm=eusm7 \font\fiveeusm=eusm5 \fi

\newfam\eufmfam \newfam\eusmfam \textfont\eufmfam=\teneufm
\scriptfont\eufmfam=\seveneufm
\scriptscriptfont\eufmfam=\fiveeufm
\textfont\eusmfam=\teneusm \scriptfont\eusmfam=\seveneusm
\scriptscriptfont\eusmfam=\fiveeusm

\def\frak{\ifmmode\let\next\frak@\else
 \def\next{\errmessage{Use \string\frak\space only in math
 mode}}\fi\next} \def\frak@#1{{\frak@@{#1}}}
 \def\frak@@#1{\fam\eufmfam#1} 
 \def\sh{\ifmmode\let\next\sh@\else
 \def\next{\errmessage{Use \string\sh\space only in math
 mode}}\fi\next} \def\sh@#1{{\sh@@{#1}}}
 \def\sh@@#1{\fam\eusmfam#1}

\ifcase\@ptsize \font\tenmsa=msam10 \font\sevenmsa=msam7
 \font\fivemsa=msam5 \font\tenmsb=msbm10
 \font\sevenmsb=msbm7 \font\fivemsb=msbm5 \or
 \font\tenmsa=msam10 scaled \magstephalf
 \font\sevenmsa=msam7 \font\fivemsa=msam5
 \font\tenmsb=msbm10 scaled \magstephalf
 \font\sevenmsb=msbm7 \font\fivemsb=msbm5 \or
 \font\tenmsa=msam10 scaled \magstep1 \font\sevenmsa=msam7
 \font\fivemsa=msam5 \font\tenmsb=msbm10 scaled \magstep1
 \font\sevenmsb=msbm7 \font\fivemsb=msbm5 \fi

\newfam\msafam \newfam\msbfam \textfont\msafam=\tenmsa
\scriptfont\msafam=\sevenmsa
\scriptscriptfont\msafam=\fivemsa \textfont\msbfam=\tenmsb
\scriptfont\msbfam=\sevenmsb
\scriptscriptfont\msbfam=\fivemsb

\def\Bbb{\ifmmode\let\next\Bbb@\else
 \def\next{\errmessage{Use \string\Bbb\space only in math
 mode}}\fi\next} \def\Bbb@#1{{\Bbb@@{#1}}}
 \def\Bbb@@#1{\fam\msbfam#1} \def\hexnumber@#1{\ifnum#1<10
 \number#1\else \ifnum#1=10 A\else\ifnum#1=11
 B\else\ifnum#1=12 C\else \ifnum#1=13 D\else\ifnum#1=14
 E\else\ifnum#1=15 F\fi\fi\fi\fi\fi\fi\fi}
 \def\msa@{\hexnumber@\msafam} \def\msb@{\hexnumber@\msbfam}
 \mathchardef\square="0\msa@03

\makeatother
\newcommand{\beq}{\begin{equation}}
\newcommand{\eeq}{\end{equation}}

\newcommand{\ba}{\begin{array}}
\newcommand{\ea}{\end{array}}
\newcommand{\bea}{\begin{eqnarray}}
\newcommand{\eea}{\end{eqnarray}}
\newcommand{\bean}{\begin{eqnarray*}}
\newcommand{\eean}{\end{eqnarray*}}

\newcommand{\be}{\begin{equation}}
\newcommand{\ee}{\end{equation}}

\newtheorem{theorem}{Theorem}[section]

\newtheorem{remark}[theorem]{Remark}

\newtheorem{proof}{Proof.}

\makeatother
 \newcommand{\RR}{{\Bbb R}}

 \def\be{\beta}

\def\be{\begin{equation}}
\def\ee{\end{equation}}

\preprint{MIT-CTP-3313 \\ \\ {\tt
hep-th/}}

\title{5d Black holes, wrapped fivebranes and 
3d Chern-Simons Super Yang-Mills}

\author{Gaetano Bertoldi
\footnote{
Research supported in part
by the CTP and the LNS of MIT and the U.S. Department of Energy 
under cooperative research agreement \# DE-FC02-94ER40818.
G. B. is also supported in part by the INFN ``Bruno Rossi''
Fellowship.}
\\
Center for Theoretical Physics,
\\ Massachusetts Institute of Technology\\ Cambridge MA 02139\\
\email{bertoldi@mit.edu}
}

\abstract{We study extremal and non-extremal generalizations 
of the regular non-abelian solution found by Chamseddine and 
Volkov in 5d N=4 gauged supergravity, which has 
been shown by Maldacena and Nastase to 
describe a system of NS5-branes wrapping an $S^3$ 
dual to 
three-dimensional $U(N)$ ${\cal N}=1$ supersymmetric Yang-Mills
with Chern-Simons coupling $k=\frac{N}{2}$.
All black hole solutions have a temperature larger 
than the Hagedorn temperature $T_c$ of the little string theory
and their entropy decreases as the temperature increases.
This is a sign that the system is thermodynamically
unstable above $T_c$. 
We have also found an analytical solution 
describing NS5-branes wrapped on a constant radius
$S^3$ and involving a linear dilaton.
Its non-extremal generalization has a temperature 
equal to $2T_c$.
} 

\begin{document}

\section{Introduction}

In \cite{Witten:1999}, 
Witten studied three-dimensional $U(N)$ ${\cal N}=1$
Super Yang-Mills with Chern-Simons coupling $k$.
He showed that this model preserves supersymmetry
for $k \geq N/2$, and he conjectured
that supersymmetry is spontaneously broken
for $k < N/2$.  
In the limit case, $k=N/2$, the Witten index is one, 
so there is one ground state and furthermore 
this vacuum is confining. 
 
Maldacena and Nastase found a supergravity dual 
of this model \cite{Maldacena:2001pb}.
In particular, they showed that  
the lifting to IIB supergravity of the regular 
solution found by Chamseddine and Volkov
in \cite{Chamseddine:2001mah}, in the context of
5d N=4 gauged supergravity,
corresponds to a system of NS5-branes wrapping a three-sphere
$S^3$ with a twisting that preserves two supercharges. 
This $S^3$ is contractible and 
the low-energy limit of this system exactly reproduces the model  
above.

This solution 
has a non-trivial flux of the NS three-form $H$ on 
the wrapped $S^3$, 
which gives rise to a Chern-Simons coupling 
equal to $N/2$.
If we add extra fivebranes wrapping the three-sphere
transverse to the original set of $N$ branes,
we increase the flux of $H$ and therefore increase 
the Chern-Simons coupling $k$.
Conversely, adding antibranes reduces the value
of $k$, but this breaks supersymmetry
as conjectured by Witten \cite{Maldacena:2001pb}.

Brane configurations realizing three-dimensional
Super Yang-Mills theories with Chern-Simons
couplings were constructed in \cite{Ohtaetal}.
Furthermore, a derivation of the supersymmetry
breaking conditions for
these models using the $s$-rule was given in   
\cite{Bergman:9908} and \cite{KOhta:1999}.

The aim of this paper is to study this three-dimensional model
at finite temperature and the
transition between the confined and the deconfined phase
in the spirit of \cite{Witten:1998}.
Our analysis will essentially follow the lines
of \cite{Gubser:2001eg}, where the case of 
4d pure ${\cal N}=1$ Super Yang-Mills
was investigated  
by considering non-extremal 
deformations of the supergravity dual proposed by Maldacena and
Nu\~nez in \cite{Maldacena:2001yy}.

\section{The supergravity description}

We briefly review the system studied in \cite{Maldacena:2001pb}.
Let us consider a system of type IIB NS5-branes 
wrapped on a three-sphere $S^3$.
The worldvolume theory will not be supersymmetric unless there is
a proper twisting \cite{BVS:1995}. This means that the spin connection
of the curved part of the worldvolume, namely $S^3$, 
has to be embedded
into the R-symmetry group of the theory, which is also 
the structure group of the normal bundle of the NS5-branes, namely
$SO(4) \cong SU(2)_L \times SU(2)_R$. 
Since the tangent space of $S^3$ is three-dimensional, 
the spin connection lies in $SU(2)$. 
Therefore, the twisting amounts to choosing an embedding of $SU(2)$
into $SO(4)$.
In particular, setting the spin connection to be in $SU(2)_L$
preserves ${\cal N}=1$ supersymmetry in three dimensions. 

If the sphere is large, then at low energies
compared to the six-dimensional coupling constant
there is a six-dimensional $U(N)$ theory
on the worldvolume of $N$ NS5-branes.
The theory will be effectively
three-dimensional at energies lower than 
the inverse radius of the sphere
and its coupling will be weak 
if $Vol(S^3) >> (\alpha')^{3/2}$.
The only massless fields are the gauge bosons and the gauginos.

In the regime where the supergravity description is
valid, the scale where the 3d theory becomes strongly coupled
and the Kaluza-Klein scale have the same order of
magnitude.

An important twist in the story is that a flux of the NS
three-form field strength $H$ on the wrapped sphere induces a 
Chern-Simons coupling in three dimensions \cite{Acharya:2000mu}.
In the S-dual description, namely where we consider $D5$-branes
wrapping $S^3$, this is a consequence of the Wess-Zumino coupling
between the RR two-form $C$ and the worldvolume gauge field strength
\cite{Douglas:1995}
$$
\frac{1}{16 \pi^3} \int_{\Sigma_6} C \wedge Tr(F \wedge F) =
-\frac{1}{16 \pi^3} \int_{\Sigma_6} G \wedge Tr(A \wedge dA +
\frac{2}{3} A^3) 
$$
\begin{equation}
= 
-\frac{k_6}{4\pi} \int_{\Sigma_3} Tr(A \wedge dA +\frac{2}{3} A^3)\,,
\label{CS1}\end{equation}
where $k_6$ is the Chern-Simons coupling appearing in the six-dimensional
Lagrangian.
It is important to stress that the effective three-dimensional
Chern-Simons coupling may in principle be different from $k_6$.
In general, integrating out massive fermions induces
a shift of this coupling whose sign
depends on the sign of the fermion mass.
In \cite{Maldacena:2001pb}, the authors showed 
that the final coupling is given by
\begin{equation}
k=k_6-\frac{N}{2}\,.
\label{kfinal}\end{equation}

\subsection{The dual gravity solution}

As explained in \cite{Maldacena:2001pb},
following the approach pioneered in \cite{Maldacena:2001mw}
and exploited in \cite{Maldacena:2001yy} to provide a
supergravity dual of pure ${\cal N}=1$ SYM in four dimensions, 
the natural setup to look for such a gravity solution
dual to the above SYM-CS theory 
would be minimal 7d gauged supergravity with gauge group $SU(2)$
\cite{TvN:1983}. 
This theory contains the metric, a dilaton $\phi$, 
the $SU(2)$ gauge fields $A^i,\, i=1,2,3$,
and a three-form field strength $h$.
A solution of this theory can then be uplifted to type IIB
using the formulas in \cite{Chamseddine:2000pl},\cite{Nastase:2000}
and \cite{Lu:2000bb}.

The seven dimensions manage to accomodate the six-dimensional brane
worldvolume, comprising the 
three-dimensional flat part and the three-sphere, 
and a radial direction.  
Most importantly, 
the $SU(2)$ gauge fields describe the 
$SU(2)_L$ within the R-symmetry group of the NS5-branes.
We can choose the large radius asymptotics of the solution
to be
$$
d s^2_{7,str} \sim d x^2_{2+1} + 
N\alpha' \left[ d r^2 + R^2(r) d \Omega_3^2 \right]\,,
$$ 
$$
\frac{1}{(2\pi)^2} \int_{S^3_{\infty}} h = k\,,
$$
$$
A^i \sim \frac{1}{2} \theta^i\,,
$$
$$
\phi \sim - r\,,
$$
where the $\theta^i$'s are the left-invariant one-forms
on the three-sphere satisfying $d \theta^i + \frac{1}{2} 
\epsilon_{ijk} \theta^j \wedge \theta^k = 0$.
The radius of $S^3$ 
will have a non-trivial $r$ dependence.
In particular, $R^2$ will vanish in the $r \to 0$ limit
whereas $R^2 \sim r$ for large $r$. 

A solution of
this $SU(2)$ seven-dimensional gauged supergravity
can be mapped to a solution of type IIB 
supergravity where we keep only a subset of the
bosonic fields, in particular the metric, the dilaton and the 
NS three-form field strength $H$ 
\cite{Chamseddine:2000pl}\cite{Nastase:2000}\cite{Lu:2000bb}
$$
d s^2_{10,str} = d s^2_{7,str} 
+ N\alpha' \frac{1}{4} \sum_{i=1}^3 (\hat \theta^i - A^i)^2
$$
\begin{equation}
H = N \left[ 
-\frac{1}{24}\epsilon_{ijk} (\hat \theta^i - A^i)
(\hat \theta^j - A^j)(\hat \theta^k - A^k) +
\frac{1}{4} F^i (\hat \theta^i - A^i) \right] + h\,,
\label{IIB}\end{equation}
where $A^i$ and $h$ are the seven-dimensional
gauge fields and three-form respectively.
Note that the ten-dimensional dilaton is the same as the 
seven-dimensional one. 
By $\hat \theta^i$'s we denote the left-invariant one-forms
on the three-sphere transverse to the branes.
Note that the transverse three-sphere is not
contractible.

Solutions to seven-dimensional supergravity describing
this system of wrapped $5$-branes 
had been considered in \cite{Acharya:2000mu}.
However, the solutions studied there develop a ``bad'' singularity
at the origin, according to the criterion of \cite{Gubser:2000nd}. 

It was realized 
in \cite{Schvellinger:2001ib} and \cite{Maldacena:2001pb} 
that a regular solution could be achieved by
uplifting the non-singular BPS solution found by
Chamseddine and Volkov in five dimensions 
\cite{Chamseddine:2001mah}.
These authors considered a proper truncation of the 
five-dimensional N=4
$SU(2) \times U(1)$ supergravity introduced by Romans
in \cite{Romans:1986ps}.
The bosonic fields are the metric, the $SU(2)$ gauge fields $A^i$,
a $U(1)$ gauge field $a$, with field strength ${\cal F}=da$,
a pair of two-forms and the dilaton.
Both these two-forms and the abelian coupling can be consistently 
set to zero on-shell.

Chamseddine and Volkov started with the following ansatz
$$
d s^2_{5,str} = -dt^2 + N\alpha' \left[ dr^2 + R(r)^2 d \Omega_3^2
\right]
$$
\begin{equation}
A^i = \frac{w(r)+1}{2} \theta^i
\label{Ai1}\end{equation}
$$
{\cal F} = Q(r)\, dt dr\,,
$$
and found two BPS solutions, both of them preserving two supercharges. 
The details of these solutions will be rederived and discussed later.

As we said, the solution can be uplifted to seven-dimensional
gauged supergravity using formulas in \cite{Cowdall:1997}
\cite{Cowdall:1998}. In particular
$$
d s^2_{7,str} = d s^2_{5,str} + d x_1^2 + d x_2^2\,,  
$$
and the $SU(2)$ gauge fields are the same.   
The abelian field strength ${\cal F}$
gives rise to a four-form field strength 
$F_{(4)}={\cal F}dx_1 dx_2$ which
can be finally dualized to yield
$h = e^{2\phi} *_{7,str} F_{(4)}$ \cite{Schvellinger:2001ib}
\cite{Maldacena:2001pb}. 

Supergravity duals of three-dimensional theories with ${\cal N}=1$
and ${\cal N}=2$ supersymmetry were also constructed in 
\cite{Hernandez:2001bh}\cite{Gomis:2001vg}\cite{Gomis:2001aa}
and \cite{Gauntlett:0110034}.
Solutions of eleven-dimensional supergravity 
corresponding to RG flows between 3d theories
with ${\cal N}=1$, ${\cal N}=2$ and ${\cal N}=3$
supersymmetry were given in \cite{Gursoy:2002}.
RG flows to 3d theories were also studied in 
\cite{Nunez:2001} in the context of 6d
gauged supergravity. 

In summary, the supersymmetric solution has the form
$\RR^{1,2} \times M_7$ with non-trivial NS flux and dilaton
and it preserves two supercharges. 
Note that the manifold $M_7$ is not Ricci flat and thus cannot be
a $G_2$-manifold. However, the analysis
of \cite{Gauntlett:0205050} shows that $M_7$ is actually
endowed with a $G_2$-structure.

\section{The type IIB supergravity description of NS-branes
on $S^3$}

We shall study solutions in the following subsector of
the type IIB supergravity action 
\begin{equation}
S_{10} = \frac{1}{4} \int d^{10}x \sqrt{-g} \left( R - \frac{1}{2}
(\partial \Phi)^2 - \frac{1}{12} e^{-\Phi} H^2   
\right)\,,
\label{IIBaction}\end{equation} 
where $H = dB = \frac{1}{6} H_{MNS} dx^M \wedge d x^N \wedge d x^S$
is the NS field strength and $\Phi$ is the dilaton.
Motivated by the previous discussion on the
uplifting of the five-dimensional solutions,
we will start from the following ansatz for the 
ten-dimensional string frame metric
\begin{align}
ds^2_{str,10} 
= &\,-e^{2X(r)} dt^2 + N \alpha' \left[ e^{2Y(r)-2X(r)}dr^2 + R^2 d\Omega_3^2 \right]
\nonumber \\
+ &\sum_{i=1}^2 dx^i dx^i +
\frac{N \alpha'}{4} \sum_{i=1}^3 (\hat \theta^i - A^i)^2\,,
\label{stringmetric}\end{align}
where $d\Omega_3^2=\sum_i \theta_i^2$ and 
the gauge fields $A^i$ are given by
(\ref{Ai1}).
The Einstein metric is given by $ds^2_E = e^{-\Phi/2} ds^2_{str}$.
The ansatz for the NS three-form field strength is given by
(\ref{IIB}) where 
\begin{equation}
h = N f(r) \frac{1}{6} \epsilon_{ijk} \theta^i \theta^j \theta^k\,,
\label{h}\end{equation}
and
\begin{equation}
F^i = d A^i + \frac{1}{2} \epsilon_{ijk} A^j \wedge A^k\,
= \frac{1}{2}w' dr \wedge \theta^i + \frac{w^2-1}{8}\,\epsilon_{ijk}
\theta^j \wedge \theta^k\,. 
\label{Fi}\end{equation}
The function $f(r)$ is determined by 5d supergravity to be
\begin{equation}
f(r) = \frac{w^3 - 3 w + 4 \kappa}{16}\,.
\label{fdierre}\end{equation}
Note that the NS three-form $H$ is closed, which,
as explained in \cite{Gauntlett:0205050},
is one of the conditions $M_7$ has to satisfy
for the existence of supersymmetric solutions. 
The constant $\kappa$ is related to the Chern-Simons parameter $k$
and the number of colors $N$. In fact, by (\ref{CS1})(\ref{kfinal})
\begin{equation}
k_6 = \frac{1}{4 \pi^2} \int_{S^3_\infty} H 
= \frac{N}{2} + N \kappa = \frac{N}{2} + k\,,
\end{equation}
where we used the fact that
$\lim_{r \to \infty} w(r) = 0$.
This implies that
\begin{equation}
\kappa = \frac{k}{N}\,.
\label{kappa}\end{equation}
We will see that the regular 
BPS solution corresponds to $\kappa = \frac{1}{2}$ and hence to
$k = \frac{N}{2}$ \cite{Maldacena:2001pb}.
Inserting the ansatz (\ref{stringmetric})--(\ref{fdierre}) 
into the type IIB action (\ref{IIBaction}), 
integrating and dropping a surface term and an overall
constant factor, we find $S_{eff} = \int dt \int dr L$, where
$$
L = \frac{1}{256} e^{-2\Phi+2X-Y} R 
\left(
16 R^2 {\Phi'}^2 - 48 R \Phi' R' + 24  {R'}^2
- 3 {w'}^2 - 16 R^2 \Phi' X' + 24 R R' X'
\right) 
$$
\begin{equation}
- \frac{1}{256 R^3} e^{-2\Phi+Y}
\left(
128 f^2 + 3 R^2(w^2-1)^2 -24 R^4 -16 R^6
\right)\,,
\label{Leff}\end{equation}
and primes denote derivatives with respect to $r$.
Note that $Y$ plays the role of a lagrangian multiplier.
It enforces a constraint that is a remnant of the residual 
invariance of the ansatz under a reparametrization of the
radial coordinate.
Varying the above one-dimensional Lagrangian
yields the following system of equations
\begin{equation}
\Phi'' = \frac{3 {w'}^2}{8 R^2} + \frac{3 R''}{2 R}\,,
\label{syst1}\end{equation}
\begin{equation}
w'' + (\frac{\nu'}{\nu} + \frac{R'}{R} - 2 \Phi')w'
-\frac{(w^2-1)(4 \kappa + (4R^2-3)w + w^3)}{2 \nu R^4} = 0\,,
\label{syst2}\end{equation}
\begin{equation}
R'' + \frac{{w'}^2 - 4 {R'}^2}{R} + \frac{4(R^2+1)}{\nu R}
+ \frac{2 \nu'}{\nu}(R \Phi' - R') - 4 R {\Phi'}^2 + 10 R' \Phi'
- \frac{(w^2-1)^2}{4 \nu R^3} =0\,,
\label{syst3}\end{equation}
\begin{equation}
\nu' = K \frac{e^{2\Phi}}{R^3}\,,
\label{syst4}\end{equation}
where $\nu \equiv e^{2X}$, $K$ is a constant
and we set $Y=0$.
This is supplemented by the constraint
$$
\frac{3 + 2 R^2}{32} - 3\frac{(w^2-1)^2}{256 R^2}
- \frac{(w^3-3w+4\kappa)^2}{512 R^4}
- \frac{1}{16}\nu R^2 {\Phi'}^2 +
\frac{3}{16}\nu \Phi' R R' 
$$
\begin{equation}
- \frac{3}{32}\nu {R'}^2
+ \frac{3}{256}\nu {w'}^2 + \frac{1}{32}\nu' R^2 \Phi'
-\frac{3}{64}\nu' R R' =0\,.
\label{syst5}\end{equation}
The above system is invariant under  
$\Phi \to \Phi + C, \, K \to e^{-2C} K$
and separately under $w \to -w, \, \kappa \to - \kappa$.
Furthermore, the system is symmetric under a
constant rescaling
of the radial coordinate. 
In fact,
if $\{\Phi(r), R(r), w(r), \nu(r)\}$ is a solution, 
then $\{\Phi(e^{2d} r) - d, R(e^{2d} r), w(e^{2d} r),
e^{-4d} \nu(e^{2d} r)\}$ is a solution as well.
The same holds for translations, by replacing $r$ with
$r + r_0$ in the argument of each function.

\section{Extremal solutions}

We can now study solutions of this system.
We can consider two distinct cases: $K=0$ or $K \ne 0$.
The first is the extremal case, where
the solution has $SO(1,2)$ symmetry. This corresponds
to $\nu = e^{2X} = const$.
By virtue of the rescaling symmetry, we can always set $\nu = 1$.
Then the ten-dimensional string frame metric reads
\begin{align}
ds^2_{str,10} =
&\,- dt^2 + N \alpha' \left[ dr^2 + R^2 d\Omega_3^2 \right]
+ \sum_{i=1}^2 dx^i dx^i +
\frac{N \alpha'}{4} \sum_{i=1}^3 (\hat \theta^i - A^i)^2\,.
\label{extremal1}\end{align}
Thus, we see that the $(t,r)$ part of the metric is flat
and the solutions will be either globally regular or have naked 
singularities.

We will also consider the non-extremal case, $K \ne 0$.
In fact we want to find black hole solutions with a regular event horizon,
since these correspond to the finite temperature gauge theory in a
deconfined phase \cite{Witten:1998}.

\subsection{BPS solutions}

The goal of this section is to find the supersymmetric 
solutions of the above system.
They correspond to the case $K=0$ and will actually satisfy  
a system of first order equations, which
was first derived in the context of $D=5$,
${\cal N}=4$ gauged supergravity in \cite{Chamseddine:2001mah}.
We will not follow the canonical approach of setting to zero
the fermion supersymmetry variations, but instead we will
try to find a superpotential $W$ for the action.
The first order system relevant to this ansatz was also
calculated by \cite{Maldacena:2001pb}. 
\noindent
The effective Lagrangian with $K=0$
$$
L = \frac{e^{-2\Phi-Y}}{256} R \left(
16 R^2 \Phi'{}^2 - 48 R \Phi' R' + 24 R'{}^2 - 3 w'{}^2
\right)
$$
$$
-\frac{e^{-2\Phi+Y}}{512 R^3}
\left[
(w^3-3w+4\kappa)^2 + 6(w^2-1)^2 R^2 - 48 R^4 - 32 R^6
\right]
$$
can be rewritten in the form
$$
L = G_{ij}(y)\frac{d y^i}{d r}\frac{d y^j}{d r} - U(y)\,,
\quad y^i = (s,g,w)\,,
$$
where $s = \Phi-\frac{3}{2}g $, $R = e^g$, and the potential $U$ 
and the metric $G_{ij}$ are
\begin{equation}
U = \frac{e^{-2s-6g+Y}}{512}
\left[
(w^3-3w+4\kappa)^2 + 6(w^2-1)^2 e^{2g} - 48 e^{4g} - 32 e^{6g}
\right]\,,
\label{U}\end{equation}
\begin{equation}
G_{ij} = \frac{e^{-2s-Y}}{16} diag\left(1,-\frac{3}{4},-\frac{3}{16}e^{-2g}
\right)\,.
\label{G}\end{equation}
A direct calculation shows that the potential $U$ can 
be represented as
$$
U = - G^{ij} \frac{\partial W}{\partial y^i}
\frac{\partial W}{\partial y^j}\,,
$$
where the superpotential $W$ reads
\begin{equation}
W = \pm\frac{3}{64} e^{-g-2s}
\sqrt{ M }\,,
\label{W}\end{equation}
with
\begin{equation}
M = \frac{4 e^{2g}}{9} - \frac{2}{3}(w^2-1)
+\frac{1}{4}e^{-2g}(w^2-1)^2 +
\left[
\frac{e^{-2g}}{24}(2w^3-6w+8\kappa)-w
\right]^2\,.
\label{M}\end{equation}
Thus the Lagrangian is equivalent to
$$
L = G_{ij}
\left(\frac{d y^i}{d r} - G^{ik}\frac{\partial W}{\partial y^k} \right)
\left(\frac{d y^j}{d r} - G^{jl}\frac{\partial W}{\partial y^l} \right)
+ 2 W'\,,
$$
which implies that the solutions to the first order equations
$$
\frac{d y^i}{d r} = G^{ik}\frac{\partial W}{\partial y^k}\,,
$$
solve the second order system as well.
We obtain
\begin{equation}
\frac{ d R}{d r} = \frac{1}{6\sqrt{M}}
\left[ 
\frac{(w^3-3w+4\kappa)^2}{8 R^4} + \frac{(w^4-8\kappa w +3)}{R^2}
+2(w^2+2)
\right]\,,
\label{BPSR}\end{equation}
\begin{equation}
\frac{d w}{d r} = \frac{4 R}{3\sqrt{M}}
\left[ 
\frac{1}{16 R^4}(w^3-3w+4\kappa)(1-w^2) 
+ \frac{1}{2 R^2}(2\kappa-w^3) - w
\right]\,,
\label{BPSw}\end{equation}
\begin{equation}
\frac{ d \Phi}{d r} = \frac{3}{2 R}\frac{d R}{d r} -\frac{3}{2}\frac{\sqrt{M}}{R}\,.
\label{BPSphi}\end{equation}
 
If $\kappa = 0$, $w \equiv 0$ solves the system and we 
retrieve the singular BPS solution found in \cite{Chamseddine:2001mah}
\begin{equation}
R = \sqrt{2r}\,, \quad \Phi = \Phi_0 - r + \frac{3}{8} \log r\,.
\label{BPSsing}\end{equation}
Note that, by (\ref{fdierre}), the abelian gauge field
is trivial.
On the other hand, the supersymmetric solution which is dual to the $D=3$
$U(N)$ ${\cal N}=1$ super Chern-Simons theory with $k=\frac{N}{2}$ 
corresponds to $\kappa = \frac{1}{2}$ \cite{Chamseddine:2001mah}. 
We can set the radius $R$ to be vanishing at $r=0$.
We will see in the next section how the requirement 
of regularity of the solution essentially 
fixes $\kappa$.
The fields have the following asymptotic behaviour
for small $r$
$$
w = 1 - \frac{1}{3} r^2 + {\cal O}(r^4) \,, \quad R = r - \frac{1}{12} r^3 + 
{\cal O}(r^5)\,,
$$
\begin{equation}
\Phi = \Phi_0 -\frac{7}{24}r^2 + {\cal O}(r^4)\,,
\label{BPSreg1}\end{equation}
whereas for large $r$
$$
R = \sqrt{2 r} + {\cal O}(e^{-2r}) \,, \quad \Phi = \Phi_0 - r + 
\frac{3}{8} \log r + {\cal O}(e^{-2r})\,,
$$
\begin{equation}
w = {\cal O}(e^{-2 r})\,.
\label{BPSreg2}\end{equation}
Note that the abelian gauge field is non-trivial
in this case.

\subsection{Non-BPS solutions}

Let us proceed to find extremal solutions of the second order system
which are not supersymmetric.

\subsubsection{Special solutions: linear dilaton \& constant radius}

There are special solutions
characterized by a constant radius
and a linear dilaton.
They are given by
\begin{equation}
(w,\kappa) =(0,0)\,, \quad R = \frac{1}{2}\,, 
\quad \Phi = \Phi_0 - Z r\,, \quad
\nu = \frac{4}{Z^2} \,.     
\label{special1}\end{equation}
Setting $Z=2$, the string frame metric reads
\begin{equation}
d s^2_{10,str} = -dt^2 + \sum_{n=1}^2 dx_n dx_n 
+ N\alpha' \left[ d r^2 
+ d M^2_6 \right]\,,
\label{special2}\end{equation}
where 
\begin{equation}
d M^2_6 = 
\frac{1}{4}\sum_{i=1}^3 (\theta^i)^2
+ \frac{1}{4} \sum_{i=1}^3 (\hat \theta^i - \frac{1}{2}
\theta^i)^2 \,,
\label{M6}\end{equation}
is a homogeneous metric on $S^3 \times S^3$.
The three-form is given by
\begin{equation}
H = N \left[ 
-\frac{1}{24}\epsilon_{ijk} 
(\hat \theta^i - \frac{1}{2}\theta^i)
(\hat \theta^j - \frac{1}{2}\theta^j)
(\hat \theta^k - \frac{1}{2}\theta^k) - \frac{1}{32}
\epsilon_{ijk} \theta^i \theta^j (\hat \theta^k - \frac{1}{2}\theta^k) 
\right]\,.
\label{Hspecial}\end{equation}
We expect the above background to be described
by a WZW model, probably a deformation
\footnote{We thank J.~Maldacena for this
suggestion.} 
of $SU(2) \times SU(2)$.
A similar solution, corresponding to
a system of NS5-branes wrapping a
constant radius $S^2$, and the
related WZW model were found in 
\cite{Gubser:2001eg}.
The description is in terms of an 
$SU(2) \times SU(2)/U(1)$ coset model studied in
\cite{PZT} and based 
on a general construction by \cite{GMM}.

We will also see that (\ref{special1})
have a simple non-extremal generalization.
It can also be checked that there are no solutions 
with constant radius and linear dilaton 
for $(w,\kappa)=(1,\frac{1}{2})$.

\subsubsection{Globally regular solutions}

In this subsection, we will consider general extremal non-BPS 
solutions of the second order system with a non-constant $w$.
For $K=0$, $\nu = 1$, the system reduces to
\begin{equation}
\Phi'' = \frac{3 {w'}^2}{8 R^2} + \frac{3 R''}{2 R}\,,
\label{systextr1}\end{equation}
\begin{equation}
w'' + (\frac{R'}{R} - 2 \Phi')w'
-\frac{(w^2-1)(4 \kappa + (4R^2-3)w + w^3)}{2 R^4} = 0\,,
\label{systextr2}\end{equation}
\begin{equation}
R'' + \frac{{w'}^2 - 4 {R'}^2}{R} + \frac{4(R^2+1)}{R}
- 4 R {\Phi'}^2 + 10 R' \Phi'
- \frac{(w^2-1)^2}{4 R^3} =0\,,
\label{systextr3}\end{equation}
$$
\frac{3 + 2 R^2}{32} - 3\frac{(w^2-1)^2}{256 R^2}
- \frac{(w^3-3w+4\kappa)^2}{512 R^4}
- \frac{1}{16}R^2 {\Phi'}^2 +
\frac{3}{16}\Phi' R R' 
$$
\begin{equation}
- \frac{3}{32}{R'}^2
+ \frac{3}{256}{w'}^2 =0\,.
\label{systextr4}\end{equation}
We are interested in globally regular solutions only,
for which spacetime is geodesically complete.
In particular, we will consider solutions with a regular origin,
which is the point $r_0$ where the three-sphere radius $R$ vanishes
but the curvature is bounded.
Note that, by the above
equations, this condition implies that $\kappa = \frac{1}{2}$
modulo the $w \to -w$, $\kappa \to -\kappa$ symmetry.

By translational symmetry, we can set $r_0 = 0$. 
We cannot analytically continue the manifold to negative $r$
and so we can assume $r \geq 0$.
Then, the system admits a one-parameter family of solutions
with the following small $r$ Taylor expansion
$$
w = 1 - b r^2 + {\cal O}(r^4)\,, \quad
R = r - \frac{2 + 9 b^2}{36} r^3 + {\cal O}(r^5)\,,
$$
\begin{equation}
\Phi = \Phi_0 - \frac{2 + 3 b^2}{8} r^2 + {\cal O}(r^4)\,.
\label{Tayreg}\end{equation}
We see that $b$ and $\Phi_0$ are free parameters.
The value
$$
b = \frac{1}{3}\,,
$$
corresponds to the regular BPS solution.
In order to find the regular non-BPS deformations,
we will numerically integrate the system 
(\ref{systextr1})-(\ref{systextr4}), using
(\ref{Tayreg}) as the boundary conditions at $r=0$.

Numerically we find that the allowed range for
$b$ is $[0, 1[$. This is due to the fact that
for $b \geq 1$, $R$ goes to zero again at a finite value of $r$. 
For $0 < b < \frac{1}{3}$, the function $w$ is always positive,
whereas for $b > \frac{1}{3}$, it has one node.
In both cases $\lim_{r \to \infty} w = 0$.
The regular BPS solution corresponds to $b = \frac{1}{3}$, 
in which case $w$ goes to zero exponentially. 

\subsection{Asymptotic behaviour of the solutions}\label{Asymptotics}

In order to evaluate the energy and  
free energy of a solution, we will need to know
explicitly its asymptotic behaviour in the limit of large $r$.

In the following, we will treat both the extremal and the non-extremal
cases at the same time and assume that for large $r$ the 
radius $R$ is not bounded.
With this assumption, we find the following
$$
R = \sqrt{2 x} - \left(\frac{\gamma^2}{4 \sqrt{2} x^{3/2}} + 
\ldots \right) + \sqrt{2} {\cal P} x^{3/4} e^{-2x} (1+ \frac{1}{x}  +\ldots)
+ {\cal O}(e^{-3x})\,,  
$$
$$
\Phi = \Phi_{\infty} - x + \frac{3}{8}\log x -
\left(\frac{15 \gamma^2}{128 x^2} 
+ \ldots
\right) 
$$
$$
+ \frac{3}{2}{\cal P} x^{1/4}e^{-2x}(1 + \frac{1}{2x}  + \ldots) +
{\cal O}(e^{-3x})\,, 
$$
$$
w = \frac{\gamma}{\sqrt{x}}\left(1 + \ldots \right)
+ {\cal C} x^{1/2} e^{-2x}(1 + \ldots) + {\cal O}(e^{-3x})\,,
$$
\begin{equation}
\nu = \frac{1}{\mu^2}
\left(1 -\frac{K}{2^{5/2}\, x^{3/4}} e^{-2x + 2\Phi_{\infty}}(1+....) 
\right)\,,
\quad x \equiv \mu (r + r_\infty)\,.
\label{asymptotics}\end{equation}
where $\mu, r_\infty, {\cal P}, \Phi_\infty, \gamma$ and ${\cal C}$
are integration constants.
Note that there is $6$ of them, due to the fact that (\ref{syst1})-(\ref{syst5}) 
can be viewed as a system of 
$7$ first order equations supplemented by one constraint,
which appeared due to the remaining reparametrization invariance
of the ansatz. The system allows
for solutions with bounded $R$ as well.
The parameter $\mu$ naturally appears due to the scaling symmetry of
the system.
The BPS solution has $\gamma_{BPS}=0$, since $w$ vanishes
exponentially, and ${\cal P}_{BPS}=0$.

\section{The non-extremal case: black hole solutions}

\subsection{Solutions with a regular horizon}

In this section, we are going to study non-extremal solutions,
where the function $\nu$ is not constant, 
corresponding to the parameter $K$ being non-vanishing. 
The ten-dimensional string frame metric reads
\begin{align}
ds^2_{str,10} 
= &\,-\nu dt^2 + N \alpha' \left[ \frac{1}{\nu} dr^2 + R^2 d\Omega_3^2 \right]
+ \sum_{i=1}^2 dx^i dx^i +
\frac{N \alpha'}{4} \sum_{i=1}^3 (\hat \theta^i - A^i)^2\,.
\label{stringmetricRH}\end{align}
Note that we can set $K=1$ without loss of generality,
since $K$ can be rescaled by
shifting the dilaton $\Phi$ by a constant.
Non-extremal solutions may have a regular event horizon 
which is our object of interest.
A solution has a regular event horizon at $r = r_h$ if 
$\nu$ has a simple zero there and all the other functions
are finite and differentiable.
Since the system is symmetric under a shift of $r$, we can also
set $r_h=0$. Then, such solutions will have the following Taylor 
expansion close to $r=0$
$$
\nu = K \frac{e^{2 \Phi_h}}{R^3_h} r  + {\cal O}(r^2)\,,
$$
$$
w = w_h + \frac{e^{-2\Phi_h}}{2 K R_h}
(w_h^2-1)(4 \kappa + (-3 + 4 R_h^2)w_h + w_h^3)r  + {\cal O}(r^2)\,,
$$
$$
\Phi = \Phi_h - \frac{e^{-2\Phi_h}}{8 K R_h^3}
\left(16 R_h^6 +3 R_h^2
(w_h^2-1)^2 + (w_h^3-3w_h+4\kappa)^2 \right)r 
+ {\cal O}(r^2)
\,,
$$
\begin{equation}
R = R_h + \frac{e^{-2\Phi_h}}{8 K R_h^2}
\left( 16 R_h^4 - 4 R_h^2
(w_h^2-1)^2- (w_h^3-3w_h+4\kappa)^2 \right) r
+ {\cal O}(r^2)\,.
\label{Taybh}\end{equation}
Here $\Phi_h, R_h$ and $w_h$, the value of the 
dilaton, the radius and $w$ at the horizon, are free parameters.
Again, we will numerically integrate (\ref{syst1})-(\ref{syst5})
towards large $r$ using (\ref{Taybh}) as initial conditions.
Note that the value of $\kappa$ is not constrained to be 
$\frac{1}{2}$ as in the case of the extremal globally 
regular solutions.
Also the set of black hole solutions is three-dimensional 
and thus has one dimension more than the set of globally regular solutions.
The extra parameter basically determines the radius of the event horizon.
In order to simplify the analysis, we will set the value of the dilaton
at the horizon to be $\Phi_h=0$.
Choosing a different value would simply amount to a rescaling
of the solutions and not affect 
their qualitative structure.

In the case of $(w_h,\kappa)=(0,0)$, we can find the solution analytically.
It is given by
\begin{equation}
(w,\kappa) =(0,0), \quad R = \frac{1}{2}, 
\quad \Phi = \Phi_0 - Z r\,, \quad
\nu = \frac{4}{Z^2} -
K \frac{4}{Z} e^{2 \Phi_0 - 2 Z r}\,.     
\label{BHanalitico}\end{equation}
Note that, setting $K=0$, we recover the 
extremal solution (\ref{special1}).
For $K \ne 0$, the $(t,r)$ part of the 10d metric
in the euclidean case corresponds to  
the ``cigar'' \cite{cigar}.   
 
In the following, we are going to consider the case $\kappa = \frac{1}{2}$ only,
since this is the value corresponding to the Chern-Simons coupling 
$k$ being equal to $\frac{N}{2}$. 
One can then see that for $w_h^2 >1$ the function
$w$ diverges. Thus, we can restrict our attention to $w_h \in [-1,1]$.
For each set of values $\{\Phi_h, R_h, w_h\}$, we will obtain a black hole
solution defined either on a finite interval or an infinite one.
In this respect the value of $R_h$ is important. 
If $R_h \gtrapprox \frac{1}{2}\sqrt{1-w_h^2}$,
then the solution extends to infinity and $R$ is asymptotic
to $\sqrt{r}$ as $r \to \infty$. 
On the other hand, if $R_h < \frac{1}{2}\sqrt{1-w_h^2}$,
then $R$ goes to zero at some finite value of $r$.

\subsection{Hawking temperature}

The ten-dimensional string frame metric reads
\begin{align}
d s^2_{str,10} = -\nu dt^2 + N \alpha' \left[ \frac{1}{\nu}
dr^2 + R^2 d\Omega_3^2 \right]
+ \sum_{i=1}^2 dx^i dx^i +\frac{N \alpha'}{4} \sum_{i=1}^3 (\hat \theta^i - A^i)^2\,.
\end{align}
Then, the $(t,r)$ part of the metric continued to the Euclidean region
is 
$$
d s^2_2 = \nu d \tau^2 + \frac{N \alpha'}{\nu}d r^2\,. 
$$
Close to the horizon $r=0$, we have $\nu \sim \nu' r$, where
$\nu' = \frac{K}{R_h^3} e^{2\Phi_h}$, and the metric is
approximately 
$d s^2 = (\nu' r d \tau^2 + \frac{N \alpha'}{\nu' r} d r^2 )$.
Introducing $\rho = \sqrt{ \frac{4 N \alpha' r}{\nu'}}$ and
$\theta = \frac{\nu'}{\sqrt{4 N \alpha'}} \tau$, the metric becomes
$d s^2 = \rho^2 d\theta^2 + d \rho^2$.
Thus, in order to have a regular metric, 
$\theta$ should be periodic with period
$2\pi$, which implies that $\tau$ has period $\beta = \frac{4 \pi}{\nu'}
\sqrt{N \alpha'}$.
The string frame metric is asymptotically flat, and 
the temperature at infinity will be given by
\begin{equation}
T = \frac{\beta^{-1}}{\sqrt{\nu(\infty)}} = 
\frac{K \,e^{2 \Phi_h}}{4 \pi R_h^3 \sqrt{N \alpha' \nu(\infty)}}\,, 
\label{Htemptrue}\end{equation} 
which takes into account the normalization
of $\nu$ at infinity.
Note that $T$ is invariant under a rescaling of $r$, which implies that
the temperature does not actually depend on $\nu(\infty)$.
Furthermore, $T$ is invariant under $K \to e^{-2C} K\,,
\Phi \to \Phi + C$. Therefore, the temperature will depend on 
three parameters only, $T=T(w_h, R_h, \kappa)$.
  
Again, we will rectrict our attention to the case $\kappa=\frac{1}{2}$,
which physically corresponds to the Chern-Simons parameter $k$
being equal to $\frac{N}{2}$.        
We found numerically that as $R_h \to \infty$  
the temperature decreases and is
asymptotic to $T_c = \frac{1}{2\pi\sqrt{N\alpha'}}$,
the Hagedorn temperature of the little string theory.
Conversely, as the black hole radius decreases the temperature increases.
In particular, for $w_h = \pm 1$, the temperature diverges in the limit
$R_h \to 0$.
This is the same behaviour found in  
\cite{Gubser:2001eg}, where the little string theory
dual to four-dimensional ${\cal N}=1$ Super Yang-Mills
was studied.

Note also that the Hawking temperature of the black-hole solution 
(\ref{BHanalitico}), corresponding to $\kappa = 0$, 
is given by $T = \frac{1}{\pi \sqrt{N \alpha'}} = 2 T_c$.

In summary, black holes exist for any value of the temperature
higher than the Hagedorn temperature $T_c$.

\section{Free energy}

Once we have obtained both the extremal and non-extremal 
generalizations of the regular BPS solution 
dual to three-dimensional super Chern-Simons theory with
$k=\frac{N}{2}$, we are ready to study their
contribution to the thermodynamics. 
In particular, we will compute the free energy of these
solutions.
 
The ten-dimensional Euclidean metric in the 
Einstein frame, with periodic time $\tau \in [0,\beta]$,
reads
\begin{equation}
d s^2 =
e^{-\Phi/2} \left( \nu d \tau^2 + N \alpha' \left[ \frac{1}{\nu}
dr^2 + R^2 d\Omega_3^2 \right]
+ \sum_{i=1}^2 dx^i dx^i +
\frac{N \alpha'}{4} \sum_{i=1}^3 (\hat \theta^i - A^i)^2 \right)\,.
\label{Euclidea1}\end{equation}
The free energy $F$ is defined by $I = \beta F$, where
$I$ is the Euclidean ten-dimensional action,
which consists of both a volume and a surface term
\begin{equation}
I = \frac{1}{4} \int_{\Omega} d^{10}x \sqrt{-g} \left(- R + \frac{1}{2}
(\partial \Phi)^2 + \frac{1}{12} e^{-\Phi} H^2   
\right) - \frac{1}{2} \int_{\Sigma} {\cal K} 
d\Sigma \equiv I_{vol} + I_{surf}\,.
\label{euclideanAction}\end{equation}

The volume integral is taken over a ten-dimensional volume 
$\Omega$ bounded by a nine-dimensional boundary $\Sigma$,
which we take to be a hypersurface at constant $r$.
The value of this constant will eventually be taken to infinity.
${\cal K}$ is the extrinsic curvature of the boundary, given by
$$
{\cal K} = \nabla_\mu N^\mu = \frac{1}{\sqrt{g}}\partial_\mu \left(
\sqrt{g} N^\mu \right)\,,
$$ 
where $N^\mu$ is the unit normal to $\Sigma$.

Then, by (\ref{Euclidea1}), $N^\mu = \sqrt{\frac{\nu}{N \alpha'}}
e^{\Phi/4} \delta^\mu_r$, and 
the metric induced on the boundary is 
$$
d s^2_b = e^{-\Phi/2} \left( \nu d \tau^2 + N \alpha' \,R^2 d\Omega_3^2 
+ \sum_{i=1}^2 dx^i dx^i +
\frac{N \alpha'}{4} \sum_{i=1}^3 (\hat \theta^i - A^i)^2 \right)
$$ 
which implies that
$d \Sigma = \frac{1}{64} \sqrt{\nu} (N \alpha')^3 e^{-9\Phi/4}R^3
\sin \theta_1 \sin \theta_4 d\tau dx_1 dx_2 d \theta_1 \ldots
d \theta_6$.

The on-shell value of the volume term of the Euclidean action,
$I_{vol}$, will reduce to the integral of a total derivative,
and can thus be expressed in terms of surface integrals.
Using the equations of motion, which is most easily done
in the string frame, we find that
$$
I_{vol} = -\frac{1}{8} \int_{\Omega} d^{10}x 
\frac{\sin \theta_1 \sin \theta_4}{64} (N \alpha')^{7/2}
\partial_r \left( 
R^3 e^{-2\Phi} \frac{\nu}{\alpha' N} \partial_r \Phi   
\right)\,
$$
\begin{equation}
= \lim_{r \to \infty} 
\frac{1}{2}\pi^4 L^2 (N \alpha')^{5/2} 
\beta\left(
-R^3 e^{-2\Phi} \nu \Phi' \right)\,,
\label{Ivolonshell}\end{equation}
where
$$
L^2 = \int dx^1 dx^2\,.
$$
Note also that the lower integration limit, $r=0$,
makes no contribution, since it either corresponds to
the origin of the coordinate system for regular solutions,
where $R=0$, or to the horizon in the case of black holes,
where $\nu=0$.

Let us now turn to the surface term, $I_{surf}$.
The extrinsic curvature is given by
$$
{\cal K} = \frac{e^{5\Phi/2}}{R^3} \partial_r \left( 
R^3 \sqrt{\frac{\nu}{N \alpha'}} e^{-9\Phi/4}
\right)\,. 
$$
Then
$$
I_{surf} = -\frac{1}{2} \int_{\Sigma} {\cal K} d\Sigma =
-\frac{1}{2} \int_{\Sigma} d^9 \tilde x 
\frac{\sin \theta_1 \sin \theta_4}{64} (N \alpha')^{5/2}
\sqrt{\nu} e^{\Phi/4} \partial_r \left(
R^3 \sqrt{\nu} e^{-9\Phi/4} 
\right) 
$$
$$
= -\frac{1}{2} \int_{\Sigma} d^9 \tilde x 
\frac{\sin \theta_1 \sin \theta_4}{64} (N \alpha')^{5/2}
\left(
3 R^2 R' \nu e^{-2\Phi} + R^3 \nu' e^{-2\Phi} /2 
-\frac{9}{4} R^3 \nu e^{-2\Phi} \Phi'  
\right) 
$$
\begin{equation}
= - \lim_{r \to \infty}
2 \pi^4 L^2 (N \alpha')^{5/2}
\beta \left(
3 R^2 R' \nu e^{-2\Phi} -\frac{9}{4} R^3 \nu e^{-2\Phi} \Phi'  
+ \frac{K}{2} 
\right)\,,
\label{Isurfonshell}\end{equation}
where we used the e.o.m. for $\nu$, namely 
$\nu' = K \frac{e^{2\Phi}}{R^3}$.
Putting everything together, we find that
\begin{equation}
I = - 2 \pi^4 L^2 (N \alpha')^{5/2} \beta
\lim_{r \to \infty} \left( 
\nu (R^3 e^{-2\Phi})' + \frac{K}{2}
\right)\,.
\label{onshellAction}\end{equation}
Therefore, the on-shell value of the action is expressed
in terms of the asymptotic values of the various fields at infinity,
which we analyzed in section \ref{Asymptotics}.

\subsection{The regularized action}

The above expression for the free energy is actually
divergent since the dilaton grows linearly as $r \to \infty$.
Hence, it needs to be regularized.
The way to do it is to subtract the value of the action for 
a reference background, and the natural choice is the regular BPS
solution.
The BPS metric is given by (\ref{Euclidea1}) with
$R=R_{BPS}$, $\Phi=\Phi_{BPS}$ and with $\nu=1$.

In order for the regularization procedure to be well-defined,
the temperature of the black hole solution should be 
matched with the temperature of the BPS solution.
To this end, we will assume that the coordinate $\tau$
has the same period $\beta$ for both solutions,
but we will modify the BPS metric by a constant factor
$\nu_{BPS}$ in the following way
$$
d s^2 = e^{-\Phi_{BPS}/2} \left( \nu_{BPS} d \tau^2 + N \alpha' \left[ 
dr^2 + R_{BPS}^2 d\Omega_3^2 \right] \right.
$$
\begin{equation}
\left. + \sum_{i=1}^2 dx^i dx^i +
\frac{N \alpha'}{4} \sum_{i=1}^3 (\hat \theta^i - A^i_{BPS})^2 \right)\,.
\label{EuclideaBPS}\end{equation}
As a result, the effective temperature of the BPS solution
is given by $\beta_{eff} = \beta \sqrt{\nu_{BPS}}$.

Repeating the calculation above, we find that
$$
I_{vol}(BPS) =
-\frac{1}{8} \int d^{10}x 
\frac{\sin \theta_1 \sin \theta_4}{64} (N \alpha')^{5/2}
\partial_r \left( R^3 e^{-2\Phi} \sqrt{\nu_{BPS}} \Phi' \right)_{BPS}\,
$$
\begin{equation}
= -\frac{1}{2} \pi^4 L^2 (N\alpha')^{5/2} \beta
\lim_{r\to\infty} 
\left( R^3 e^{-2\Phi} \sqrt{\nu_{BPS}} \Phi' \right)_{BPS}\,.
\label{IvolBPS}\end{equation}
Since the unit normal to the boundary is now 
$N^\mu = \frac{1}{\sqrt{N \alpha'}} e^{\Phi/4} \delta^\mu_r$,
we find that
\begin{equation}
I_{surf}(BPS) = 
-2 \pi^4 L^2 (N \alpha')^{5/2}
\beta \lim_{r \to \infty} \sqrt{\nu_{BPS}} \,
\left( 3 R^2 R' e^{-2\Phi} - \frac{9}{4}
R^3 \Phi' e^{-2\Phi} 
\right)_{BPS}\,.
\end{equation}
Finally, the regularized action $I_{reg} \equiv I - I_{BPS}$ reads
\begin{equation}
I_{reg} = 
- 2 \pi^4 L^2 (N \alpha')^{5/2} \beta 
\lim_{r \to \infty} \left(
\nu (R^3 e^{-2\Phi})' - \sqrt{\nu_{BPS}}\,
(R^3 e^{-2\Phi})_{BPS}' - \frac{K}{2}
\right)\,.
\label{Iregularized}\end{equation}
The free energy is then defined by
\begin{equation}
F \equiv \frac{I_{reg}}{\beta} =
-2\pi^4 L^2 (N\alpha')^{5/2}
\lim_{r \to \infty} \left(
\nu (R^3 e^{-2\Phi})' 
- \sqrt{\nu_{BPS}} (R^3 e^{-2\Phi})'_{BPS} + \frac{K}{2}
\right)\,.
\label{freeenergy}\end{equation}
Again, in order to take this limit 
in a sensible way, we need to impose proper matching conditions
at the boundary $\Sigma$ \cite{HH}.
First of all, the geometries induced on $\Sigma$ must
be the same in both backgrounds.
Since $\Sigma \cong S^1 \times S^3 \times T^2 \times S^3_{transverse}$,
the geometries will be the same if the following conditions
are satisfied on $\Sigma$
\begin{equation}
e^{-\Phi/2} \nu = e^{-\Phi_{BPS}/2} \nu_{BPS}\,,
\quad e^{-\Phi/2} R^2 = e^{-\Phi_{BPS}/2}R^2_{BPS}\,,
\quad \Phi = \Phi_{BPS}\,,
\label{Mcond1}\end{equation}
and $w$ converges to $w_{BPS}$ sufficiently fast.

\subsection{Energy and Entropy}\label{Energyentropy}

In \cite{HH}, Hawking and Horowitz showed that
for stationary spacetimes admitting foliations by
spacelike hypersurfaces $\Sigma_t$, the regularized
free energy, obtained from the action as we did above,
is related to the energy via the usual thermodynamic
equation
\begin{equation}
F = E - T S\,,
\label{termo1}\end{equation}
where $T = \beta^{-1}$, $S$ is the entropy, and $E$ is the 
conserved ADM energy defined by 
\begin{equation}
E = -\frac{1}{2} \int_{S_t^\infty} \sqrt{|g_{00}|}\,\left( 
{}^8 K - {}^8 K_0
\right) dS^\infty_t\,.
\label{ADM}\end{equation}
The integration is carried out over the $8$-dimensional boundary of
the $9$-dimensional hypersurface $\Sigma_t$ and
${}^8 K$ and ${}^8 K_0$ are the extrinsic curvatures of
$S_t^\infty$ in the geometry under study and in the reference background
geometry respectively. 
The two $8$-dimensional geometries on $S_t^\infty$ must be the same,
and it is also assumed that the $g_{00}$ components are equal at
$S_t^\infty$.
Finally, it is also required that the matter fields at the boundary 
agree at least up to a sufficiently high order \cite{HH}.
As in \cite{Gubser:2001eg}, the results of this analysis can be 
applied to our case.

Let us use (\ref{ADM}) to calculate the energy of our solutions.
The metric induced on the constant time hypersurface $\Sigma_t$
reads
$$
d s_t^2 = 
e^{-\Phi/2} \left( N \alpha' \left[ \frac{1}{\nu}
dr^2 + R^2 d\Omega_3^2 \right]
+ \sum_{i=1}^2 dx^i dx^i +
\frac{N \alpha'}{4} \sum_{i=1}^3 (\hat \theta^i - A^i)^2 \right)\,.
$$
Conversely the BPS reference solution will induce the following metric
$$
d s^2_{t,BPS} = 
e^{-\Phi_{BPS}/2} \left( N \alpha' \left[
dr^2 + R^2_{BPS} d\Omega_3^2 \right]
+ \sum_{i=1}^2 dx^i dx^i +
\frac{N \alpha'}{4} \sum_{i=1}^3 (\hat \theta^i - A_{BPS}^i)^2 \right)\,.
$$
The boundary $S^\infty_t$ of $\Sigma_t$ is defined by the
hypersurface at constant $r$ in the limit where $r$ goes to infinity.
The boundary is then given topologically by the product of
two three-spheres, namely the three-sphere wrapped by the
NS-branes and $S^3_{trasnverse}$, 
and the two-torus $T^2$ with coordinates $x_1$ and $x_2$.

The geometries induced on $S_t^\infty$ will be the same if and only if
\begin{equation}
e^{-\Phi/2} R^2 = e^{-\Phi_{BPS}/2} R^2_{BPS}\,, 
\quad \Phi = \Phi_{BPS}\,, 
\label{MatchEnergy}\end{equation}
and $w$ converges to $w_{BPS}$ sufficiently fast.
The $g_{00}$ components of the two backgrounds agree if
$$
e^{-\Phi/2} \nu = e^{-\Phi_{BPS}/2} \nu_{BPS}\,.
$$
Note that these conditions are the same as (\ref{Mcond1})
required in the evaluation of the regularized action.
The unit normal to $S_t^\infty$ is given by 
$n^k = \sqrt{\frac{\nu}{N \alpha'}} e^{\Phi/4} \delta^k_r$, 
so that 
$$
{}^8 K = \frac{\sqrt{\nu}}{R^3}
e^{+9\Phi/4} \partial_r \left( 
\frac{1}{\sqrt{N \alpha'}} R^3 e^{-2\Phi}
\right)\,.
$$
Conversely, $n^k_{BPS} = 
\frac{1}{\sqrt{N \alpha'}} e^{\Phi_{BPS}/4} \delta^k_r$,
and ${}^8 K_0 = \frac{1}{R^3}
e^{+9\Phi_{BPS}/4} \partial_r \left( 
\frac{1}{\sqrt{N \alpha'}} R^3 e^{-2\Phi_{BPS}}
\right)$.
Finally
\begin{equation}
E = - 2 \pi^4 L^2 (N \alpha')^{5/2}
\lim_{r\to\infty}
\left(
\nu \left( R^3 e^{-2\Phi} \right)' - \sqrt{\nu_{BPS}} \,
\left( R^3_{BPS} e^{-2\Phi_{BPS}} \right)'
\right)\,.
\label{energy}\end{equation}
This reproduces exactly the first term in (\ref{freeenergy}),
which agrees with the general thermodynamic relation
(\ref{termo1}), and yields the following
expression for the entropy of the solutions
\begin{equation}
S = 4 \pi^5 R_h^3 L^2 (N \alpha')^3 e^{-2\Phi_h}
= \pi^4 L^2 (N \alpha')^{5/2} \,\beta K\,,
\label{entropy}\end{equation}
where we used the expression for the Hawking temperature,
(\ref{Htemptrue}), and
set $\nu(\infty)=1$. We see that the entropy is proportional to 
the geometrical area of the horizon.
Note also that although both the energy and the action are
invariant under a translation of $r$, they both 
get a factor $e^{-2C}$ under $\Phi \to \Phi+C$,
$K \to e^{-2C}K$. 

It is important to notice that for generic values of $K$ and $\Phi_h$
a black hole solution will have $\nu(\infty) \ne 1$.
Thus, in order to achieve proper normalization at infinity and
keep the value of $\Phi_h$ fixed, we need to fine-tune $K$.
Recall that the value of the dilaton at the horizon is related to
the Yang-Mills coupling constant. Hence, to compare
two different black hole solutions in a physically
meaningful way, we have to make sure that they have 
the same value of $\Phi_h$.
The fine-tuning proceeds as follows. 
First, we use the scaling symmetry under $r \to e^{2d}r$ 
to set $\nu(\infty)=1$ by taking $d = \frac{1}{4} \ln\,\nu(\infty)$.
This changes $\Phi_h$ to $\Phi_h - d$ and leaves $K$
invariant. Then, we use the symmetry $\Phi \to \Phi + C$,
$K \to e^{-2C} K$, to set $\Phi$ to a prescribed value.
This last step has no effect on $\nu(\infty)$.  

Numerical analysis shows that as the temperature decreases,
namely as $R_h$ becomes larger, the black hole 
entropy actually increases.
This is a signal of thermodynamic instability of the system
above $T_c$. 
 
We can now calculate the energy and free energy of a general solution
using the expressions found above.
Let us take a non-BPS solution and set $r_{\infty}=0$ 
in (\ref{asymptotics}).
The regular BPS solutions make up a two-parameter family,
the parameters being $\Phi_0$ and $r_\star$, which accounts
for the symmetry under translations of $r$.
These two parameters together with $\nu_{BPS}$
will be fine-tuned so that the matching
conditions (\ref{Mcond1}) are satisfied at the boundary
$\Sigma$. However, the functions $w$ and $w_{BPS}$ 
will not match exactly, unless the boundary is strictly
at infinity, in which case both functions vanish.
The discrepancy $\Delta w = w - w_{BPS}$ 
should tend to zero fast enough, otherwise the energy 
will be infinite.
A direct calculation shows that a polynomial fall-off, 
$\gamma \ne 0$, is not enough for the energy to be finite.
On the other hand, if the parameter $\gamma$ is vanishing
$\Delta w \sim e^{-2r}$ and the energy is actually finite.

\subsection{Solutions with finite energy}

Let us carry out the explicit computation of the energy
for solutions with $\gamma=0$.
The asymptotics for large $r$ are given by 
$$
R = \sqrt{2 r} + \sqrt{2} {\cal P} r^{3/4} e^{-2r} (1+ \frac{1}{r} +\ldots)\,,
\quad
\nu = 1 -\frac{K}{2^{5/2}\, r^{3/4}} e^{-2r + 2\Phi_{\infty}} + \ldots
$$
\begin{equation}
\Phi = \Phi_{\infty} - r + \frac{3}{8}\log r + 
\frac{3}{2}{\cal P} r^{1/4}e^{-2r}(1 + \frac{1}{2r}  + \ldots)\,,
\label{c1}\end{equation}
where we set $r_{\infty}=0$ and $\mu = 1$.
The asymptotics of the BPS solution are given by 
$$
R_{BPS} = \sqrt{2(r+r_*)} + \ldots, \quad \Phi_{BPS} = \Phi_*
 - (r + r_*) + \frac{3}{8} \log (r+r_*) + \ldots
$$
\begin{equation}
\nu_{BPS} = const\,.
\label{c2}\end{equation}
We need to evaluate the limit
(\ref{energy}) under the matching conditions (\ref{Mcond1}),
which are equivalent to 
\begin{equation}
\nu = \nu_{BPS}\,,
\quad e^{-2\Phi} R^3 = e^{-2\Phi_{BPS}}R^3_{BPS}\,,
\quad R = R_{BPS}\,,
\label{c3}\end{equation}
By the first of the above conditions, the limit
becomes
\begin{equation}
\lim_{r \to \infty} \sqrt{\nu} \left(
\sqrt{\nu} (R^3 e^{-2\Phi})' - (R^3 e^{-2\Phi})'_{BPS}
\right)\,.
\label{delta1}\end{equation}
Since
$$
R^3 e^{-2\Phi} = 2 \sqrt{2} r^{3/4} e^{-2 \Phi_{\infty} + 2r}
+ 3 \sqrt{2} {\cal P} e^{-2 \Phi_{\infty}} + \ldots
$$
\begin{equation}
R^3_{BPS} e^{-2\Phi_{BPS}} =
2\sqrt{2} (r + r_*)^{3/4} e^{-2 \Phi_* + 2(r + r_*)} + \ldots\,,
\label{R1}\end{equation}
we then obtain
$$
\lim_{r \to \infty} 
\sqrt{\nu} \left\{
\left(1 - \frac{K}{2^{7/2} r^{3/4}} e^{-2r+2\Phi_{\infty}} \right)
\left(4 \sqrt{2} r^{3/4} e^{-2\Phi_{\infty}+2r} 
+ \frac{3 \sqrt{2}}{2 r^{1/4}} e^{-2\Phi_{\infty}+2r} \right) 
\right.
$$
$$
\left. -
\left( 4 \sqrt{2} (r+r_*)^{3/4} e^{-2\Phi_{*}+2(r+r_*)} 
+ \frac{3 \sqrt{2}}{2 (r+r_*)^{1/4}} e^{-2\Phi_{*}+2(r+r_*)}               
\right)
\right\}
$$
$$
= \lim_{r \to \infty}
\sqrt{\nu} \left\{ 
4 \sqrt{2} r^{3/4} e^{-2\Phi_{\infty}+2r} 
- 4 \sqrt{2} (r+r_*)^{3/4} e^{-2\Phi_{*}+2(r+r_*)} 
\right.
$$
$$
\left. + \frac{3 \sqrt{2}}{2 r^{1/4}} e^{-2\Phi_{\infty}+2r}
- \frac{3 \sqrt{2}}{2 (r+r_*)^{1/4}} e^{-2\Phi_{*}+2(r+r_*)}
\right\}
- \frac{K}{2}\,.
$$
By the second condition in (\ref{c3}) and (\ref{R1})
we find that 
$$
4 \sqrt{2} (r+r_*)^{3/4} e^{-2\Phi_{*}+2(r+r_*)} =
4 \sqrt{2} r^{3/4} e^{-2\Phi_{\infty}+2r} +
6 \sqrt{2} {\cal P} e^{-2 \Phi_{\infty}} \,.
$$
Furthermore, the third condition in (\ref{c3}) 
$$
\sqrt{2 r} (1 + {\cal P} r^{1/4} e^{-2 r}) = \sqrt{2(r + r_*)}
$$
yields
$$
r_* = 2 {\cal P} r^{5/4} e^{-2 r}\,.
$$
Then, the limit becomes
$$
\lim_{r \to \infty} \sqrt{\nu} \left\{
\frac{3}{\sqrt{2}} e^{-2\Phi_{\infty}+2r} \left(
\frac{1}{r^{1/4}} - \frac{r^{3/4}}{r+r_*}
\right)
\right\} - 6\sqrt{2} {\cal P} e^{-2\Phi_{\infty}} - \frac{K}{2}
$$
$$
= \frac{6}{\sqrt{2}} {\cal P} e^{-2\Phi_{\infty} } 
- 6\sqrt{2} {\cal P} e^{-2\Phi_{\infty}} - \frac{K}{2} =
- 3 \sqrt{2} {\cal P} e^{-2\Phi_{\infty}} - \frac{K}{2}\,.
$$

Finally, by (\ref{energy}), the ADM energy of a
non-BPS solution with $\gamma=0$
will be given by
\begin{equation}
E = 2 \pi^4 L^2 (N \alpha')^{5/2}
\left(
3\sqrt{2} {\cal P} e^{-2\Phi_\infty} + \frac{K}{2}
\right)\,.
\label{energyX}\end{equation}
Note that the energy is invariant under constant shifts of $r$ and
is therefore independent of $r_{\infty}$.
Conversely, under $\Phi \to \Phi + C$, 
${\cal P}$ is invariant while $K \to e^{-2C}K$,
and the energy picks up an overall factor $e^{-2C}$.

Let us calculate the action via
$$
I = \beta E - S\,.
$$
For the globally regular solutions, the entropy vanishes and $K=0$
that gives
\begin{equation}
I_{global} = 2 \pi^4 L^2 (N \alpha')^{5/2}
\left(
3\sqrt{2}\, \beta\, {\cal P}\, e^{-2\Phi_\infty} 
\right)\,.
\label{c4}\end{equation} 
For black holes, the entropy $S$ and the term proportional to $K$ in
the expression for the energy cancel out and we find
\begin{equation}
I_{BH} = 
2 \pi^4 L^2 (N \alpha')^{3}
(3\sqrt{2}) \frac{4 \pi}{K} {\cal P} R_h^3 
e^{-2\Phi_h-2\Phi_\infty} \,,
\label{c5}\end{equation}
where we used the expression for the Hawking temperature 
(\ref{Htemptrue}) 
and normalized the solution to set $\nu(\infty)=1$.
Again, under $\Phi \to \Phi + C$, ${\cal P}$ and $R_h$
remain invariant while $K \to e^{-2C}K$, so that
the action picks up the overall factor $e^{-2C}$.

\subsection{Globally regular solutions with finite energy}

The numerical analysis of the system (\ref{systextr1})-(\ref{systextr4}) 
shows that there are actually no finite energy globally regular
solutions besides the regular BPS one.

Figure 1 is a plot of $\gamma$ as a function of $b$, which
parametrizes the family of globally regular solutions and belongs
to the interval $[0,1[$.
We see that $\gamma$ has a single zero, whose position is 
compatible with the supersymmetric solution, which corresponds to
$b=\frac{1}{3}$ (Fig. 2).  

\vspace{0.5cm}
\begin{center}
\epsfig{file=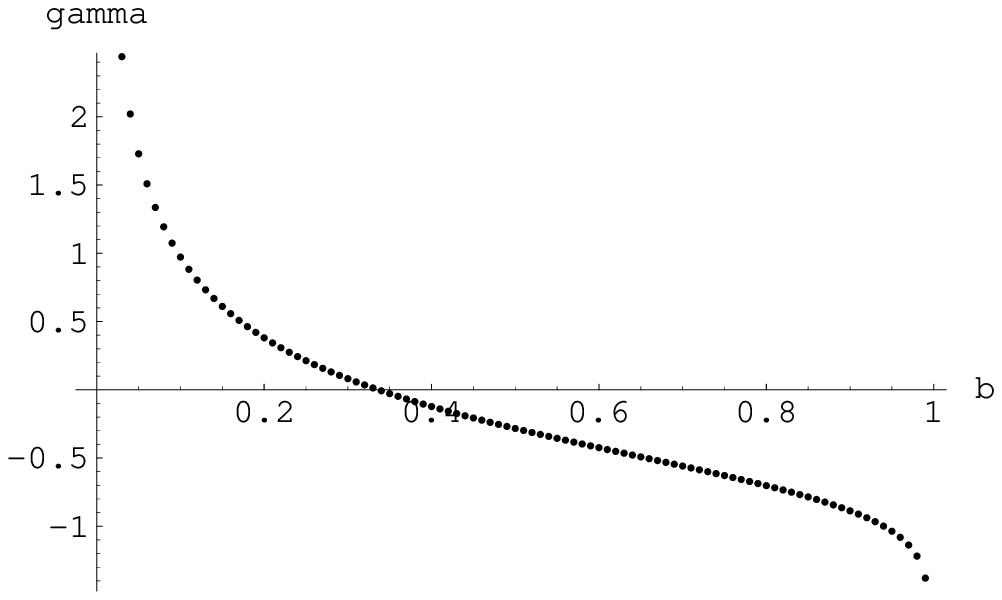, width=9cm}
\end{center}
\vspace{0.5cm}
\baselineskip=13pt
\centerline{\small{Figure 1. Globally regular solutions:}}
\centerline{\small{plot of $\gamma$ as a function of $b$.}}

\vspace{0.5cm}
\begin{center}
\epsfig{file=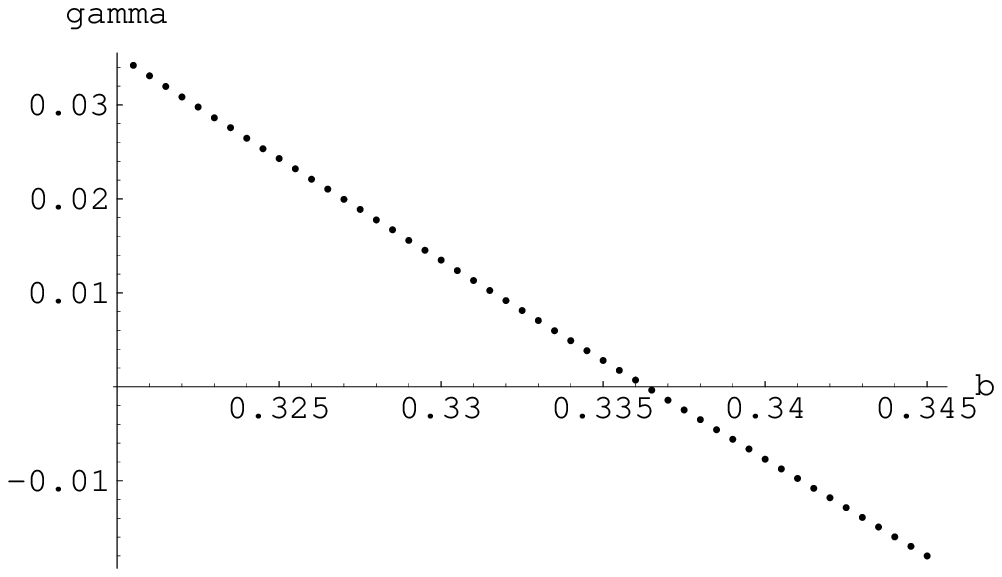, width=9cm}
\end{center}
\vspace{0.5cm}
\baselineskip=13pt
\centerline{\small{Figure 2.}}
\centerline{\small{Plot of $\gamma$ expanded around the node.}}

\vspace{0.5cm}

\baselineskip=15.5pt

\subsection{Black hole solutions with finite energy}

In the case of black-holes, a similar analysis 
shows that there is in fact a single finite energy
solution for each value of the temperature above
$T_c = \frac{1}{2\pi \sqrt{N\alpha'}}$.

In Figs. 3, 4 and 6, we show a plot of $\gamma$ as a function of the
parameter $w_h \in \,]-1,1[$ for a fixed value of $R_h$
and a constant value of the temperature $T$.

\vspace{0.5cm}
\begin{center}
\epsfig{file=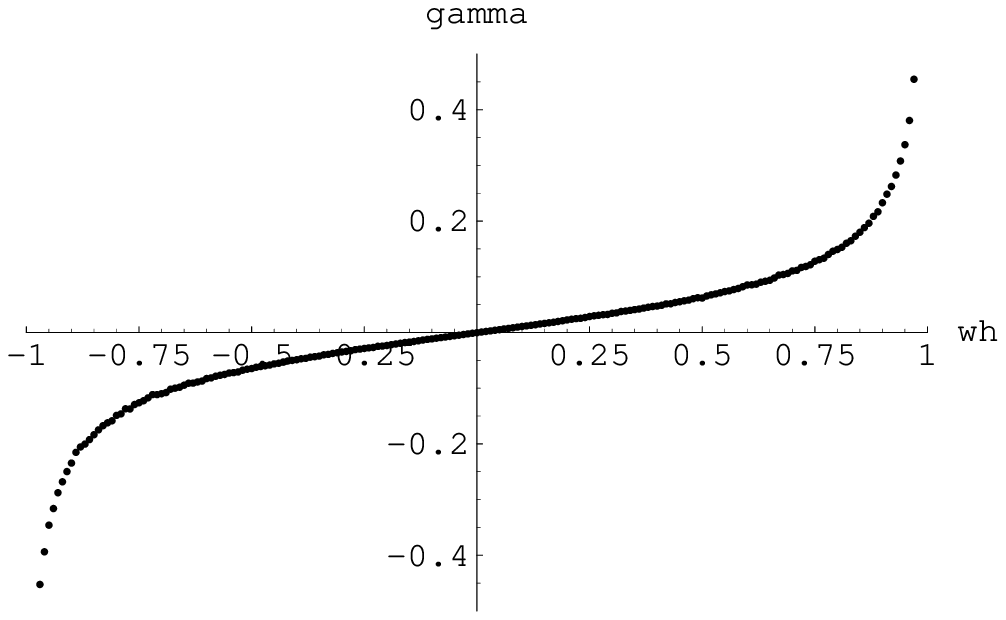, width=9cm}
\end{center}
\vspace{0.5cm}
\baselineskip=13pt
\centerline{\small{Figure 3: plot of $\gamma$ as a function of 
$w_h$ for $R_h=50$ and $T/T_c$ around $1.00$. }}


\vspace{0.5cm}
\begin{center}
\epsfig{file=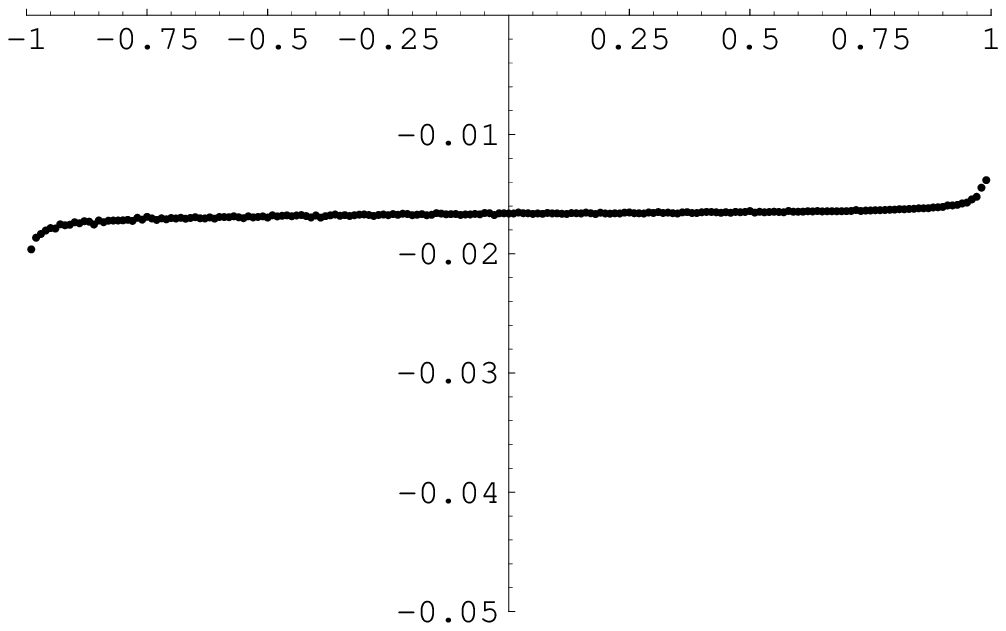, width=9cm}
\end{center}
\vspace{0.5cm}
\baselineskip=13pt
\centerline{\small{Figure 4: plot of $\gamma$ as a function of 
$w_h$ for $R_h=10$ and $T/T_c$ around $1.01$. }}



\begin{center}
\epsfig{file=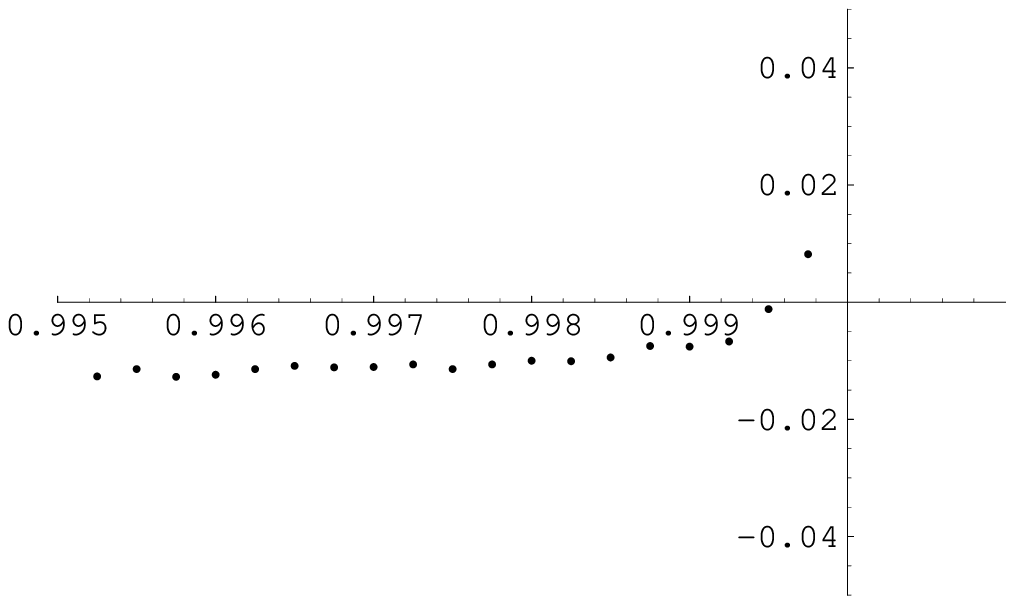, width=9cm}
\end{center}
\vspace{0.5cm}
\baselineskip=13pt
\centerline{\small{Figure 5: Still $R_h=10$ and $T/T_c$ around $1.01$. }}
\centerline{\small{Expanded version of Fig.4 to show
the node which is very close to $w_h = 1.0$}}



\vspace{0.5cm}
\begin{center}
\epsfig{file=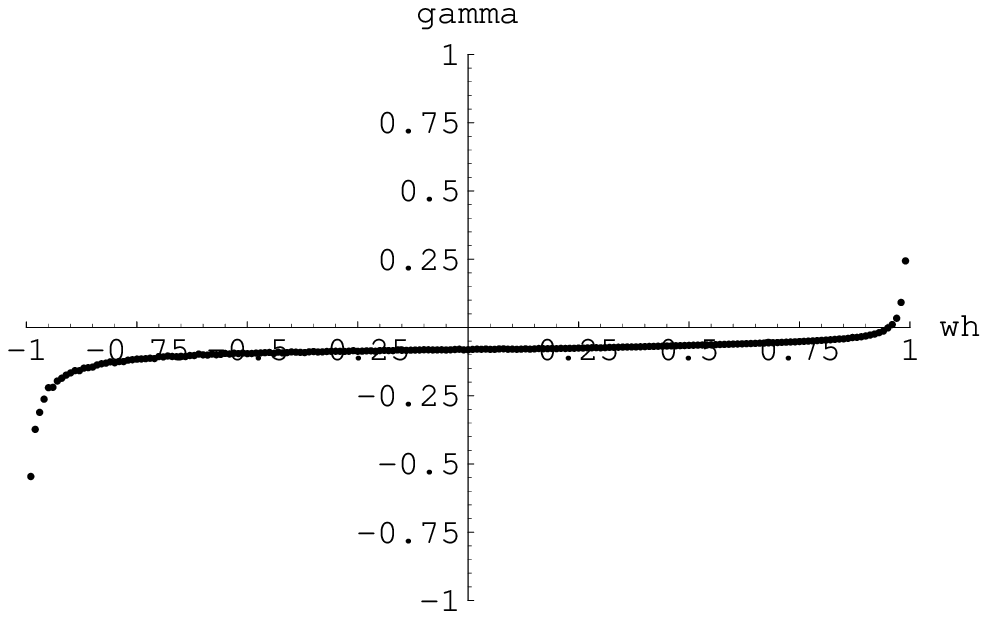, width=9cm}
\end{center}
\vspace{0.5cm}
\baselineskip=13pt
\centerline{\small{Figure 6: plot of $\gamma$ as a function of 
$w_h$ for $R_h=2$ and $T/T_c$ around $1.12$. }}
\centerline{\small{}}

The existence of finite energy black holes
may allow a Hawking-Page transition
above $T_c$ \cite{HPage}.
Below $T_c$, there are no black holes,
the BPS solution with periodic
Euclidean time dominates the path integral 
and the field theory is confining.
On the contrary, above $T_c$, if a given black hole
had a negative action, then its contribution to the path
integral would be more relevant than the BPS one
and the dual field theory would undergo a transition to
a deconfined phase. 
By (\ref{c5}), in order to detect such a transition 
one has to determine ${\cal P}$.
This is quite difficult 
since ${\cal P}$ is a coefficient in front of 
subleading terms which vanish exponentially
and we have not been able to estimate it
with accuracy. 

However, as we remarked in section \ref{Energyentropy}\,,
the fact that the black hole
entropy (\ref{entropy}) decreases as the temperature increases
tells us that 
the would be high temperature phase 
is actually unstable. 
It would have negative specific heat. 
Therefore, such a transition between the stable
low-temperature confining phase and the unstable 
high-temperature deconfining phase may not take place.
This was first pointed out in \cite{Gubser:2001eg}, 
where a similar system of NS5-branes
wrapping a shrinking $S^2$, which is dual to 
4d ${\cal N}=1$ SYM \cite{Maldacena:2001yy},
was studied in detail. In this case, 
the authors actually determined that the black hole 
action becomes negative above $T_c$\,.
A similar analysis was carried out
earlier in \cite{Buchel:200107}, where it was 
shown that a 
system of NS5-branes wrapping a two-sphere 
in a resolved conifold had negative specific heat.

The underlying little string theory is
believed to have an exponential growth in
the number of states around $T_c$ and
is actually thermodynamically unstable above this temperature 
\cite{Kutasov:200012}.
The results of \cite{Buchel:200107}
\cite{Gubser:2001eg} and the present work
confirm the presence of this instability.

This thermodynamic instability 
is thought to be due to a 
Gregory-Laflamme instability 
\cite{GLF1}\cite{GLF2}
of the underlying system of black NS5-branes. 
Based on results found in the context of 
the $AdS_4$ Reissner-Nordstrom solution, 
Gubser and Mitra conjectured that 
{\it for a black brane with translational
symmetry, a Gregory-Laflamme instability 
exists precisely when the brane is 
thermodynamically unstable}
\cite{GubserMitra1}\cite{GubserMitra2}.
Further arguments were given in \cite{Reall:2000}. 
Along these lines, Rangamani \cite{Rangamani}
argued that the
instability of the little string theory
found in \cite{Kutasov:200012} is 
actually due to the presence of
a {\it threshold unstable mode} \cite{Reall:2000} 
that survives the non-decoupling limit of
the non-extremal NS5-branes.   
It would be interesting to see whether such a 
mode exists in the present context.

\section{Conclusions}

We have studied both extremal and non-extremal 
generalizations
of the regular supersymmetric solution dual
to 3d $U(N)$ ${\cal N}=1$ Super Yang-Mills
with Chern-Simons coupling $k=N/2$, originally
found by Chamseddine and Volkov
in the context of $5d$ $N=4$ gauged 
supergravity. 
We rederived both the singular BPS solution (\ref{BPSsing})
and the regular one (\ref{BPSreg1})(\ref{BPSreg2}).

We have found an interesting analytical 
non-supersymmetric solution
corresponding to a system of NS5-branes 
wrapping a constant radius $S^3$ (\ref{special1}).
This geometry factorizes into $S^3 \times S^3$
times a non-compact part with linear dilaton
(\ref{special2})(\ref{M6}).
It has a non-trivial $H$ field (\ref{Hspecial})
and admits a non-extremal generalization
with a Hawking temperature $T=2 T_c$\,,
where $T_c$ is the Hagedorn temperature of the
little string theory (\ref{BHanalitico}).

We have also found that there are no finite energy extremal
globally regular solutions except the BPS one.
Furthermore, all black hole solutions
have a temperature which is larger than $T_c$
and their entropy decreases as the temperature increases.
This indicates that the system is thermodynamically 
unstable above $T_c$ 
as was also found in \cite{Buchel:200107}\cite{Gubser:2001eg}.

\section{Acknowledgements}

I am particularly indebted to C.~N\'u\~nez and M. Schvellinger
for initial participation in the project and 
constant support. I would also like to thank A. Hanany,
J. Maldacena, D. Renner and J. Troost for discussions and useful comments.
My work is supported in part by 
the CTP and the LNS of MIT, by the U.S. Department of Energy
under cooperative research agreement \# DE-FC02-94ER40818,
and by the INFN ``Bruno Rossi''Fellowship.


\bibliographystyle{JHEP}

\begin{thebibliography}{99}



\bibitem{Witten:1999}
E.~Witten,``Supersymmetric index of three-dimensional gauge theory'',
in *Shifman, M.A. (ed.): The many faces of the superworld* 156-184,
hep-th/9903005.


\bibitem{Maldacena:2001pb}
J.~Maldacena and H.~Nastase,
``The supergravity dual of a theory with dynamical supersymmetry  breaking'',
JHEP {\bf 0109}, 024 (2001),
hep-th/0105049.


\bibitem{Chamseddine:2001mah}
A.~H.~Chamseddine and M.~S.~Volkov,
``Non-Abelian vacua in D=5, N = 4 super gauged supergravity'',
JHEP {\bf 0104} 023 (2001),
hep-th/0101202.



\bibitem{Ohtaetal}
T.~Kitao, K.~Ohta and N.~Ohta,
``Three-dimensional gauge dynamics from brane configurations 
with (p,q)-five-brane'', Nucl.Phys.{\bf B539} 
(1999) 79, hep-th/9808111.


\bibitem{Bergman:9908}
O.~Bergman, A.~Hanany, A.~Karch and B.~Kol, 
``Branes and supersymmetry breaking in 
3D gauge theories'', JHEP {\bf9910}, 036 (1999),
hep-th/9908075.


\bibitem{KOhta:1999} 
K.~Ohta, ''Supersymmetric Index and s-rule for Type IIB Branes'',
JHEP {\bf9910} 006 (1999),
hep-th/9908120.


\bibitem{Witten:1998}
E.~Witten,
``Anti-de Sitter space, thermal phase transition, and confinement in gauge
theories'', Adv.Theor.Math.Phys. 2 (1998) 505-532,
hep-th/9803131.


\bibitem{Gubser:2001eg}
S.~S.~Gubser, A.~A.~Tseytlin and M.~S.~Volkov,
``Non-Abelian 4-d black holes, wrapped 5-branes, and their dual descriptions,'
JHEP {\bf 0109}, 017 (2001),hep-th/0108205.


\bibitem{Maldacena:2001yy}
J.~M.~Maldacena and C.~N\'u\~nez,
``Towards the large $N$ limit of pure ${\cal N} = 1$ super Yang-Mills,''
Phys.\ Rev.\ Lett.\ {\bf 86} (2001) 588, hep-th/0008001.


\bibitem{BVS:1995}
M. Bershadsky, C. Vafa and V. Sadov, 
``D-Branes and Topological Field Theories'', Nucl.Phys.{\bf B463}, 
420 (1996), hep-th/9511222.


\bibitem{Acharya:2000mu}
B.~S.~Acharya, J.~P.~Gauntlett and N.~Kim,
``Fivebranes wrapped on associative three-cycles'',
Phys.Rev.D {\bf63} (2001),
hep-th/0011190.


\bibitem{Douglas:1995}
M.R. Douglas, ``Branes within branes'', hep-th/9512077.


\bibitem{Maldacena:2001mw}
J.~Maldacena and C.~N\'u\~nez,
``Supergravity description of field theories on curved manifolds and a no go theorem,''
Int.\ J.\ Mod.\ Phys.\ A {\bf 16}, 822 (2001), hep-th/0007018.


\bibitem{TvN:1983} P.K. Townsend and P. van Nieuwenhuizen,
``Gauged Seven-Dimensional Supergravity'', Phys.Lett.B {\bf 125},
41 (1983).


\bibitem{Chamseddine:2000pl}
A.H. Chamseddine and W.A. Sabra, ``D=7 SU(2) gauged supergravity from
D=10 supergravity'', Phys.Lett.B {\bf 476}, 415 (2000), hep-th/9911180. 


\bibitem{Nastase:2000}
H. Nastase and D. Vaman, ``On the non-linear KK reduction on spheres 
of supergravity theories'', Nucl.Phys.B{\bf 583}, 211 (2000), hep-th/0002028. 


\bibitem{Lu:2000bb}
M. Cveti\^c, H.~L\"u and C.~N.~Pope,
``Consistent Kaluza-Klein Sphere Reductions,''
Phys.Rev. D{\bf62} (2000) 064028,
hep-th/0003286.


\bibitem{Gubser:2000nd}
S.~S.~Gubser,
``Curvature singularities: The good, the bad, and the naked'',
Adv.Theor.Math.Phys. 4 (2002) 679-745,
hep-th/0002160.


\bibitem{Schvellinger:2001ib}
M.~Schvellinger and T.~A.~Tran,
``Supergravity duals of gauge field theories from SU(2) x U(1) gauged  supergravity in five dimensions,''
JHEP {\bf 0106}, 025 (2001), hep-th/0105019.


\bibitem{Romans:1986ps}
L.~J.~Romans,
``Gauged ${\cal N}=4$ Supergravities In Five-Dimensions And Their 
Magnetovac Backgrounds,''
Nucl.\ Phys.\ B {\bf 267}, 433 (1986).


\bibitem{Cowdall:1997}
P.M. Cowdall, ``Supersymmetric Electrovacs in gauged supergravities'',
Class, Quant. Grav. {\bf 15}, 2937 (1998), hep-th/9710214.


\bibitem{Cowdall:1998}
P.M. Cowdall, ``On gauged maximal supergravity in six dimensions'',
JHEP {\bf 9906}, 019 (1999), hep-th/9810041.


\bibitem{Hernandez:2001bh}
R.~Hernandez,
``Branes wrapped on coassociative cycles,''
Phys.Lett. B{\bf521} (2001) 371-375,
hep-th/0106055. 


\bibitem{Gomis:2001vg}
J.~Gomis and T.~Mateos,
``D6 branes wrapping Kaehler four-cycles,''
Phys.Lett. B{\bf524} (2002) 170-176,
hep-th/0108080. 


\bibitem{Gomis:2001aa}
J.~Gomis and J.~G.~Russo,
``D = 2+1 N = 2 Yang-Mills theory from wrapped branes,''
JHEP {\bf 0110} 028 (2001), 
hep-th/0109177.


\bibitem{Gauntlett:0110034}
J.~P.~Gauntlett, N.~Kim, D.~Martelli and D.~Waldram,
``Fivebranes Wrapped on SLAG Three-Cycles 
and Related Geometry'', 
JHEP {\bf 0111} 018 (2001), hep-th/0110034. 


\bibitem{Gursoy:2002}
U.~Gursoy, C.~N\'u\~nez and M.~Schvellinger,
``RG flows from Spin(7), CY 4-fold and HK manifolds
to AdS, Penrose limits and pp waves,''
JHEP {\bf 0206} 015 (2002), 
hep-th/0203124.


\bibitem{Nunez:2001}
C.~N\'u\~nez, I.~Y.~Park, M.~Schvellinger
and T.~A.~Tran, ``Supergravity
duals of gauge theories from F(4)
gauged supergravity in six dimensions'',
JHEP {\bf 0104} 025 (2001),
hep-th/0103080.


\bibitem{Gauntlett:0205050}
J.~P.~Gauntlett, D.~Martelli, S.~Pakis and D.~Waldram,
``G-Structures and Wrapped NS5-Branes'', 
hep-th/0205050. 


\bibitem{PZT}
L.A.~Pando~Zayas and A.A.Tseytlin,
``Conformal sigma models for a class
of $T^{(p,q)}$ spaces'', Class. Quant. Grav. 17,
5125-5131, (2000), hep-th/0007086. 

\bibitem{GMM}
E.~Guadagnini, M.~Martellini and M.~Mintchev,
``Scale invariant sigma models on homogeneous spaces'',
Phys. Lett. B{\bf 194}: 69 (1987);
E.~Guadagnini, ``Current Algebra In Sigma Models On
Homogeneous Spaces'', Nucl. Phys. B{\bf290}, 417 (1987);


\bibitem{cigar}
S.~Elitzur, A.~Forge and E.~Rabinovici, ``Some global
aspects of string compactifications'',  Nucl. Phys. B{\bf359}, 581 (1991);
G.~Mandal, A.M.~Sengupta and S.R.~Wadia, ``Classical solutions of 
two-dimensional string theory'', Mod.Phys.Lett. A{\bf 6}, 1685 (1991);
E.~Witten, ``On string theory and black holes'', Phys.Rev. D{\bf 44},
314 (1991). 


\bibitem{HH}
S.W.~Hawking and G.T.~Horowitz,
``The Gravitational Hamiltonian, action, entropy and surface terms,''
Class. Quant. Grav. 13, 1487 (1996), gr-qc/9501014. 

\bibitem{HPage}
S.W.~Hawking and D.N.~Page,''Thermodynamics Of Black Holes
In Anti-De Sitter Space,'' Comm.Math.Phys. {\bf 87}, 577 (1983). 

\bibitem{Kutasov:200012}
D.~Kutasov and D.A.~Sahakyan, 
``Comments on the Thermodynamics
of Little String Theory'',
JHEP {\bf0102} 021 (2001), hep-th/0012258


\bibitem{Buchel:200107}
A.~Buchel, 
``On the thermodynamic instability of LST'', hep-th/0107102.


\bibitem{GLF1}
R.~Gregory and R.~Laflamme, ``Black strings and 
p-branes are unstable'', Phys. Rev. Lett.{\bf 70} (1993) 2837, 
hep-th/9301052. 

\bibitem{GLF2}
R.~Gregory and R.~Laflamme, ``The instability of
charged black strings and p-branes'', Nuc.Phys.{\bf B428} (1994) 399-434, 
hep-th/9404071. 

\bibitem{GubserMitra1}
S.S.~Gubser and I.~Mitra,''Instability of
charged black holes in anti-de Sitter space,''
hep-th/0009126.

\bibitem{GubserMitra2}
S.S.~Gubser and I.~Mitra,''The evolution
of unstable black holes in anti-de Sitter space,''
JHEP {\bf 0108} 018 (2001),
hep-th/0011127.

\bibitem{Reall:2000}
H.S.~Reall,''Classical and thermodynamic stability
of black branes,'' Phys.Rev. D{\bf 64} (2001) 044005,
hep-th/0104071.

\bibitem{Rangamani}
M.~Rangamani, ``Little string thermodynamics'', 
JHEP {\bf 0106} 042 (2001),
hep-th/0104125.


\end{thebibliography}

\end{document}